\documentclass[12pt]{article}
\usepackage{latexsym}
\usepackage{amsfonts}
\usepackage{epsfig}

\catcode`\@=11

\def\section{\@startsection{section}{1}{\z@}{3.5ex plus 1ex minus
 .2ex}{2.3ex plus .2ex}{\bf}}

\def\thesubsection{\arabic{section}.\arabic{subsection}}
\renewcommand{\subsection}[1]{\addtocounter{subsection}{1}
\vspace{2.5mm}\par\noindent {\it \thesubsection . #1}\par
 \vspace{0.5mm} }

\catcode`\@=12
\newfont{\mbm}{msbm10 scaled\magstep1}

\mathchardef\varGamma="0100 \mathchardef\varDelta="0101
\mathchardef\varTheta="0102 \mathchardef\varLambda="0103
\mathchardef\varXi="0104 \mathchardef\varPi="0105
\mathchardef\varSigma="0106 \mathchardef\varUpsilon="0107
\mathchardef\varPhi="0108 \mathchardef\varPsi="0109
\mathchardef\varOmega="010A

\DeclareFontFamily{U}{rsf}{}
\DeclareFontShape{U}{rsf}{m}{n}{
  <5> <6> rsfs5 <7> <8> <9> rsfs7 <10-> rsfs10}{}
\DeclareMathAlphabet\Scr{U}{rsf}{m}{n}

\begin{document}
\begin{titlepage}
\rightline{{CERN-TH/2003-295}}
\rightline{{UCLA/03/TEP/39}}
\rightline{{hep-th/0312019}}
\vskip 2cm
\centerline{{\large\bf $K3\times T^2/\mathbb{Z}_2$ orientifolds with 
fluxes,}}
\vskip 0.3cm
\centerline{{\large\bf  open string moduli and critical points}}
\vskip 1cm
\centerline{C. Angelantonj$^\dagger$,  R. D'Auria $^{\star\flat}$, S.
Ferrara $^{\dagger\ddag\sharp}$ and M. Trigiante $^\star $}
\vskip 0.5cm
\centerline{\it ${}^\dagger$ CERN, Theory Division, CH 1211 Geneva 23,
Switzerland}
\vskip 0.3cm 
\centerline{\it ${}^\ddagger$ INFN, Laboratori Nazionali di
Frascati, Italy}
\vskip 0.3cm
\centerline{\it ${}^\sharp$ University of California, Los Angeles, USA}
\vskip 0.3cm
\centerline{\it ${}^\star$ Dipartimento di Fisica, Politecnico di Torino}
\centerline{\it C.so Duca degli Abruzzi, 24, I-10129 Torino, Italy}
\vskip 0.3cm
\centerline{\it ${}^\flat$ 
Istituto Nazionale di Fisica Nucleare, Sezione di Torino,Italy}
\vskip  1.0cm
\begin{abstract}
We extend the four--dimensional gauged
supergravity analysis of type IIB vacua on $K3\times
T^2/\mathbb{Z}_2$ to the case where also D3 and D7 moduli,
belonging to ${\Scr N}=2$ vector multiplets, are turned on. 
In this case, the
overall special geometry does not correspond to a symmetric space,
unless D3 or D7 moduli are switched off. In the
presence of non--vanishing fluxes, we discuss supersymmetric
critical points which correspond to Minkowski vacua, finding
agreement with previous analysis. Finally, we point out that care 
is needed in the choice of the symplectic holomorphic sections of 
special geometry which enter the computation of the scalar potential.
\end{abstract}
\end{titlepage}

\section{Introduction}

In the present note we generalise the four--dimensional
supergravity analysis of \cite{ADFL} to the case where D3 and
D7 brane moduli are turned on, together with 3--form fluxes.
This problem seems of particular physical relevance in view of
recent work on inflationary models 
\cite{lateKallosh}--\cite{KTW}, 
whose underlying scalar
potential is in part predicted by the mechanism of supergravity
breaking, which is at work in some string compactifications in
the presence of fluxes \cite{PS}--\cite{Becker:2003yv}.

From the point of view of the four--dimensional ${\Scr N}=2$
effective supergravity, open string moduli, corresponding to
D7 and D3--brane positions along $T^2$, form an enlargement of
the vector multiplet moduli--space which is locally described, in
absence of open--string moduli, by \cite{ADFL}:
\begin{equation}
\label{stu}
    \left(\frac{{\rm SU} (1,1)}{{\rm U}(1)}\right)_s\times
\left(\frac{{\rm SU}(1,1)}{{\rm U}(1)}\right)_t\times
    \left(\frac{{\rm SU}(1,1)}{{\rm U}(1)}\right)_u\,,
\end{equation}
where $s,\,t,\,u$ denote the scalars of the vector multiplets
containing the $K3$--volume and the R--R $K3$--volume--form, the
$T^2$--complex structure, and the IIB axion--dilaton system,
respectively:
\begin{eqnarray}
s&=& C_{(4)} -{\rm i}\, {\rm Vol} (K_3),\,
\nonumber\\
t&=& \frac{g_{12}}{g_{22}} +{\rm i}\,\frac{\sqrt{{\rm det}
g}}{g_{22}}\,,
\nonumber\\
u&=& C_{(0)} +{\rm i}\, e^{\phi}\,,
\end{eqnarray}
where the matrix $g$ denotes the metric on $T^2$.

When D7--branes moduli are turned on, what is
known is that ${\rm SU}(1,1)_s$ acts as an electric--magnetic duality
transformation \cite{GZ} both on the bulk and D7--brane vector
field--strengths, while the ${\rm SU} (1,1)_u$ acts as an
electric--magnetic duality transformation on the D3--vector
field--strengths. Likewise the bulk vectors transform
perturbatively under ${\rm SU}(1,1)_u \times {\rm SU}(1,1)_t$ while the
D3--brane vectors do not transform under ${\rm SU}(1,1)_s \times
{\rm SU}(1,1)_t$ and the D7--brane vectors do not transform under
${\rm SU}(1,1)_u \times {\rm SU}(1,1)_t$. 

All this is achieved
starting from the following trilinear prepotential of special
geometry:
\begin{equation}\label{prepot}
    {\Scr F}(s,t,u,x^k,y^r)\,=\, stu-\frac{1}{2}\,s \,x^k
    x^k-\frac{1}{2}\,u\,
    y^r y^r\,,
\end{equation}
where $x^k$ and $y^r$ are the positions of the D7 and
D3--branes along $T^2$ respectively, $k=1,\dots, n_7$,
$r=1,\dots ,n_3$, and summation over repeated indices is
understood. This prepotential is unique in order to preserve the
shift--symmetries of the $s,\,t,\,u$ bulk complex fields up to
terms which only depend on $x$ and $y$.

The above prepotential
gives the correct answer if we set either all the $x^k$ or all the
$y^r$ to zero. In this case the special geometry describes a
symmetric space:
\begin{eqnarray}\label{symanifold7}
\left(\frac{{\rm SU}(1,1)}{{\rm U}(1)}\right)_s\times
\frac{{\rm SO}(2,2+n_7)}{{\rm SO}(2)\times
{\rm SO}(2+n_7)}\,,\,\,\,\,\,\,\mbox{for
$y^r=0$}\,,\\\label{symanifold3}
\left(\frac{{\rm SU}(1,1)}{{\rm U}(1)}\right)_u\times
\frac{{\rm SO}(2,2+n_3)}{{\rm SO}(2)\times
{\rm SO}(2+n_3)}\,,\,\,\,\,\,\,\mbox{for $x^k=0$}\,.
\end{eqnarray}
For both $x$ and $y$ non--vanishing, the complete K\"ahler
 manifold (of complex dimension $3+n_3+n_7$) is no longer a
 symmetric space even if it still has $3+n_3+n_7$ shift
 symmetries\footnote{The prepotential in eq. (\ref{prepot}) actually 
corresponds to the homogeneous not symmetric spaces called
$L(0,n_7,n_3)$ in \cite{van}. We thank A. van Proeyen for a
discussion on this point.}. 

 Note that for $x^k=0$ the manifold is predicted as a truncation of
 the manifold describing the moduli--space of $T^6/\mathbb{Z}_2$ 
${\Scr N}=4$ orientifold
 in the presence of D3--branes. The corresponding symplectic
 embedding was given in \cite{DFGVT}. For $y^r=0$ the
 moduli--space is predicted by the way ${\rm SU}(1,1)_s$ acts on both
 bulk and D7 vector fields. Upon compactification of Type IIB theory 
on $T^2$, the D7--brane moduli are insensitive to the further
 $K3$ compactification and thus their gravity coupling must be the same as
 for vector multiplets coupled to supergravity in $D=8$. 
Indeed if $2+n$ vector multiplets
 are coupled to ${\Scr N}=2$ supergravity in $D=8$, their 
non--linear $\sigma $--model
 is \cite{SS},\cite{ADF}:
\begin{equation}
\frac{{\rm SO}(2,2+n)}{{\rm SO}(2)\times {\rm SO}(2+n)}\times \mathbb{R}^+\,.
\end{equation}
Here $\mathbb{R}^+$ denotes the volume of $T^2$ and the other
part is the second factor in (\ref{symanifold7}). Note that in
$D=8,\,{\Scr N}=2$ the R--symmetry is ${\rm U}(1)$ which is the
${\rm U}(1)$ part of the $D=4,\,{\Scr N}=2$ ${\rm U}(2)$ R--symmetry. The
above considerations prove eq. (\ref{symanifold7}).
 
 Particular care is needed  \cite{AFT} when the effective supergravity is
 extended to include gauge couplings, as a result of turning on
 fluxes in the IIB compactification \cite{PS}.

 The reason is that the scalar potential depends explicitly on the
 symplectic embedding of the holomorphic sections of special
 geometry, while the K\"ahler potential, being symplectic
 invariant, does not. In fact, even in the analysis without open
 string moduli \cite{ADFL}, it was crucial to consider a Calabi--Visentini
 basis where the ${\rm SO}(2,2)$ linearly acting symmetry on the bulk
 fields was ${\rm SU}(1,1)_u\times {\rm SU}(1,1)_t$ \cite{ABCDFFM},\cite{CDFV}.

 In the case at hand, the choice of symplectic basis is the one
 which corresponds to the Calabi--Visentini basis for $y^r=0$,
 with the ${\rm SU}(1,1)_s $ acting as an electric--magnetic duality
 transformation \cite{ADFL}, but it is not such basis for the D3--branes
 even if the $x^k=0$. Indeed, for $x^k=0$, we must reproduce the
 mixed basis used for the $T^6/\mathbb{Z}_2$ orientifold 
\cite{FP},\cite{KST} in
 the presence of D3--branes found in \cite{DFGVT}.

 We note in this respect, that the choice of the symplectic
 section made in \cite{KTW} does not determine type IIB vacua
 with the 3--form fluxes turned on. It does not correspond in fact
 to the symplectic embedding discussed in \cite{ADFL},
 \cite{DFGVT} and \cite{AFT}. The problem arises already in 
the absence of branes.
 Indeed in \cite{KTW} the type IIB
 duality group ${\rm SU}(1,1)_u$, which is associated in their notation
 to the modulus $S$, has a non--perturbative action on the bulk
 vector fields while this action should be perturbative. As a
 consequence of this, in \cite{KTW} a potential was found
that does not  stabilise the axion--dilaton
 field, which  is in contradiction with known results 
\cite{TT}\footnote{Actually, in a revised version of \cite{KTW},
agreement with our results is found}.

\section{${\Scr N}=2$ and ${\Scr N}=1$ supersymmetric cases.}

\subsection{${\Scr N}=2$ gauged supergravity}

We consider  the gauging of ${\Scr N}=2$ supergravity with a
special geometry given by eq. (\ref{prepot}). Let us briefly
recall the main formulae of special K\"ahler geometry. The
geometry of the manifold is encoded in the holomorphic section
$\varOmega=(X^\varLambda,\,F_\varSigma)$ which, in the {\it special
coordinate} symplectic frame, is expressed in terms of a
prepotential ${\Scr F}(s,t,u,x^k,y^r)=F(X^\varLambda)/(X^0)^2={\Scr
F}(X^\varLambda/X^0)$, as follows:
\begin{equation}
\varOmega = ( X^\varLambda,\,F_\varLambda=\partial F/\partial
X^\varLambda)\,.\label{specialcoordinate}
\end{equation}
 In our case ${\Scr F}$ is given by eq. (\ref{prepot}). The
K\"ahler potential $K$ is given by the symplectic invariant
expression:
\begin{equation}
K = -\log \left[{\rm i}(\overline{X}^\varLambda
F_\varLambda-\overline{F}_\varLambda X^\varLambda)\right] \,.
\end{equation}
In terms of $K$ the metric has the form
$g_{i\bar{\jmath}}=\partial_i\partial_{\bar{\jmath}}K$. The
matrices $U^{\varLambda\varSigma}$ and $\overline{{\Scr
N}}_{\varLambda\varSigma}$ are respectively given by:
\begin{eqnarray}
U^{\varLambda\varSigma}&=&e^K\, {\Scr D}_i X^\varLambda {\Scr
D}_{\bar{\jmath}}
\overline{X}^\varSigma\,g^{i\bar{\jmath}}= -\frac{1}{2}\,{\rm Im}({\Scr
N})^{-1}-e^K\,\overline{X}^\varLambda X^\varSigma\,,\nonumber\\
\overline{{\Scr N}}_{\varLambda\varSigma}&=& \hat{h}_{\varLambda|I}\circ
(\hat{f}^{-1})^I{}_\varSigma\,,\,\,\mbox{where}\,\,\,\,
\hat{f}_{I}^\varLambda = \left(\matrix{{\Scr D}_i X^\varLambda \cr
\overline{X}^\varLambda }\right)\,;\,\,\,\,\hat{h}_{\varLambda|
I}=\left(\matrix{{\Scr D}_i F_\varLambda \cr \overline{F}_\varLambda
}\right) \,.
\end{eqnarray}
 For our choice of ${\Scr F}$, $K$ has the following form:
\begin{equation}
K = -\log\left[-8\,({\rm Im}(s)\,{\rm Im}(t){\rm
Im}(u)-\frac{1}{2}\,{\rm Im}(s)\,({\rm
Im}(x)^i\,)^2-\frac{1}{2}\,{\rm Im}(u)\,({\rm
Im}(y)^r\,)^2)\right] \,,
\end{equation}
 with ${\rm Im}(s)<0$ and ${\rm Im}(t),\,{\rm
Im}(u)>0$ at $x^k=y^r=0$. The components $X^\varLambda,\,F_\varSigma$ of
the symplectic section which correctly describe our problem, are
chosen by performing a constant symplectic change of basis from
the one in (\ref{specialcoordinate}) given in terms of the
prepotential in eq. (\ref{prepot}). The symplectic matrix is
\begin{equation}
\left(\matrix{A & -B\cr B & A}\right) \,,
\end{equation}
with 
\begin{eqnarray}
&&A\,=\,\frac{1}{\sqrt{2}}\,\left(\matrix{1 &0&0&0&0&0\cr 0
&0&-1&-1&0&0\cr -1 &0&0&0&0&0\cr 0 &0&1&-1&0&0\cr 0
&0&0&0&\sqrt{2}&0\cr 0 &0&0&0&0&\sqrt{2}}\right)\,,\nonumber\\&&
B\,=\,\frac{1}{\sqrt{2}}\,\left(\matrix{0 &1&0&0&0&0\cr 0
&0&0&0&0&0\cr 0 &1&0&0&0&0\cr 0 &0&0&0&0&0\cr 0 &0&0&0&0&0\cr 0
&0&0&0&0&0}\right)\,.
\end{eqnarray}  
The rotated symplectic sections then become
\begin{eqnarray}
X^0 &=& \frac{1}{{\sqrt{2}}}\,(1 - t\,u +
\frac{(x^k)^2}{2})\,,
\nonumber \\
X^1 &=& -\frac{t +
u}{{\sqrt{2}}}\,,
\nonumber \\
X^2 &=&  -\frac{1}{{\sqrt{2}}}\,({1 + t\,u - \frac{(x^k)^2}{2}})\,,
\nonumber\\
X^3 &=& \frac{t - u}{{\sqrt{2}}}\,,
\nonumber \\
X^k &=& x^k\,,
\nonumber \\
X^a &=& y^r\,,
\nonumber\\
F_0 &=& \frac{s\,\left( 2 - 2\,t\,u + (x^k)^2 \right) +
u\,(y^r)^2}{2\,{\sqrt{2}}}\,,
\nonumber \\
F_1 &=&
  \frac{-2\,s\,\left( t + u \right)  +
  (y^r)^2}{2\,{\sqrt{2}}}\,,
\nonumber\\
F_2&=&
  \frac{s\,\left( 2 + 2\,t\,u - (x^k)^2 \right)  -
  u\, (y^r)^2}{2\,{\sqrt{2}}}\,,
\nonumber \\
 F_3 &=& \frac{2\,s\,\left( -t + u \right)  + (y^r)^2}{{2\,\sqrt{2}}}\,,
\nonumber\\
F_i &=& -
  s\,x^k
 \,,
\nonumber \\
F_a &=& -u\,y^r\, .
\end{eqnarray}
Note that, since $\partial X^\varLambda /\partial s =0$ the new sections do 
not admit a prepotential, and the no--go theorem on partial supersymmetry
breaking \cite{Cecotti} does not apply in this case.
As in \cite{ADFL}, we limit ourselves to gauge shift--symmetries of
the quaternionic manifold of the $K3$ moduli--space. Other
gaugings which include the gauge group on the brane will be
considered elsewhere.

\subsection{${\Scr N}=2$ supersymmetric critical points}

In the sequel we limit our analysis to critical points in flat
space. The ${\Scr N}=2$ critical points  demand ${\Scr
P}^x_\varLambda=0$. This equation does not depend on the special
geometry and its solution is the same as in \cite{ADFL}, i.e.
$g_2,\,g_3\neq 0$, $g_0=g_1=0$ and $e^m_a=0$ for $a=1,2$, were the
Killing vectors gauged by the fields $A^2_\mu$ and $A^3_\mu$ are
constants and their non--vanishing components are $k^u_2=g_2$
along the direction $q^u=C^{a=1}$  and $k^u_3=g_3$ along the
direction $q^u=C^{a=2}$. The 22 fields $C^m,\,C^a$, $m=1,2,3$ and
$a=1,\dots ,19$ denote the Peccei--Quinn scalars. The vanishing of
the hyperino--variation further demands:
\begin{equation}
k^u_\varLambda\, X^\varLambda = 0 \,\,\Rightarrow
\,\,\,X^2=X^3=0\,\,\,\Leftrightarrow\,\,\,t=u \,,\,\,\,
1+t^2=\frac{(x^k)^2}{2}\,.\label{n2tu}
\end{equation}
Hence for ${\Scr N}=2$ vacua the D7 and D3--brane positions
are still moduli while the axion--dilaton and $T^2$ complex
structure are stabilised.

\subsection{${\Scr N}=1$ supersymmetric critical points}

The ${\Scr N}=1$ critical points in flat space studied in
\cite{ADFL} were first obtained by setting $g_0,\,g_1\neq 0$ and
$g_2=g_3=0$, with $k^u_0=g_0$ along the direction $q^u=C^{m=1}$
and $k^u_1=g_1$ along the direction $q^u=C^{m=2}$.

\paragraph{Constant Killing spinors.}
By imposing $\delta_{\epsilon_2}\,f=0$ for the variations of the fermionic
fields $f$  we get the following: \\
From the hyperino variations:
\begin{eqnarray}
\delta_{\epsilon_2}\,\xi^{A
a}&=&0\,\,\,\Rightarrow\,\,\,\,e^a_m=0\,\,m=1,2\,;\,\,a=1,\dots,
19\nonumber\\
\delta_{\epsilon_2}\,\xi^{A
}&=&0\,\,\,\Rightarrow\,\,\,\mbox{vanishing of the gravitino
variation}
\end{eqnarray}
The gravitino variation vanishes if:
\begin{equation}
S_{22} = -g_0\, X^0+{\rm i}\,g_1\, X^1
\end{equation}
From the gaugino variations we obtain:
\begin{equation}
\delta_{\epsilon_2}\,(\lambda^{\bar{\imath}})_A = 0\,\,\Rightarrow
\,\,e^{\frac{K}{2}}\,{\Scr P}^x_\varLambda\,(\partial_i
X^\varLambda+(\partial_i K)\, X^\varLambda)\,\sigma^x_{A2}\,=\,0\,,
\end{equation}
the second term (with $\partial_i K$) gives a contribution
proportional to the gravitino variation while the first term, for
$i=u,\,t,\,x^k$ respectively gives:
\begin{eqnarray}
-g_0\, \partial_u X^0+{\rm i}\, \partial_u X^1 &=&0 \,,\nonumber\\
-g_0\, \partial_t X^0+{\rm i}\, \partial_t X^1 &=&0 \,,\nonumber\\
-g_0\, \partial_{x^k} X^0 &=&0 \,, \nonumber\\
\end{eqnarray}
for $i=y^r$ the equation is identically satisfied. From the last
equation we get $x^k=0$ and the other two, together with
$S_{22}=0$ give $u=t={\rm i}$, $g_0=g_1$.

So we see that for
${\Scr N}=1$ vacua the D7--brane coordinates are frozen while
the D3--brane coordinates remain moduli. This agrees with the
analysis of \cite{TT}. If $g_0\neq g_1$ the above solutions give
critical points with vanishing cosmological constant but with no
supersymmetry left.

More general ${\Scr N}=1,0$ vacua can be
obtained
also in this case by setting $g_2,\,g_3\neq 0$. The only
extra conditions coming from the gaugino variations for ${\Scr
N}=1$ vacua is that $e^{a=1,2}_m=0$. This eliminates from the
spectrum two extra metric scalars $e^{a=1,2}_3$ and the
$C_{a=1,2}$ axions. These critical points preserve ${\Scr N}=1$ or
not depending on whether $|g_0|=|g_1|$ or not.

We can describe the ${\Scr N} =1 \to {\Scr N}=0$ transition with an
${\Scr N} =1$ no--scale supergravity \cite{noscale1, noscale2} based 
on a constant 
superpotential and a non--linear sigma--model which is
\begin{equation}
{{\rm U} (1,1+n_3) \over {\rm U}(1)\times {\rm U} (1+ n_3)} \times 
{{\rm SO} (2, 18) \over {\rm SO} (2) \times {\rm SO} (18)} \,,
\end{equation}
where the two factors come from vector multiplets and hypermultiplets,
respectively.
This model has vanishing scalar potential, reflecting the fact that there 
are not further scalars becoming massive in this transition \cite{ADFL}. 
We further note that
any superpotential $W(y)$ for the D3 brane coordinates would generate a 
potential \cite{FPorr} term
\begin{equation}
e^{K} \, K^{y\bar y}\, \partial_y W \partial_{\bar y} \bar W\,,
\end{equation}
which then would require the extra condition $\partial_y W =0$ for
a critical point with vanishing vacuum energy.

The residual moduli
space of K3 metrics at fixed volume is locally given by 
\begin{equation}
{{\rm SO} (1,17) \over {\rm SO} (17)}\,.
\end{equation}
We again remark that we have considered vacua with vanishing vacuum energy.
We do not consider here the possibility of other vacua with
non--zero vacuum energy, as i.e. in \cite{KTW}.

\section{The potential}

The general form of the ${\Scr N}=2$ scalar potential is:
\begin{equation}
V = 4\, e^K  h_{uv} k^u_\varLambda k^v_\varSigma
\,X^\varLambda\,\overline{X}^\varSigma+e^K g_{i\bar{\jmath}}\,
k^i_\varLambda k^{\bar{\jmath}}_\varSigma \,X^\varLambda\,
\overline{X}^\varSigma+e^K (U^{\varLambda\varSigma}-3\, e^K \,X^\varLambda
\overline{X}^\varSigma){\Scr P}^x_\varLambda\,{\Scr P}^x_\varSigma
\end{equation}
where the second term is vanishing for abelian gaugings. Here
$h_{uv}$ is the quaternionic metric and $k_\varLambda^u$ the
quaternionic Killing vector of the hypermultiplet
$\sigma$--model.

The scalar potential, at the extremum of the
$e^a_m$ scalars, has the following form\footnote{Note that there is 
a misprint in eq. (5.1) of ref. \cite{ADFL}. The term $e^{2\phi} \, e^{\tilde
K} \, g_0 \, g_1 ( X_0 \bar X_1 + X_1 \bar X_0 )$ is actually absent}:
\begin{eqnarray} 
\label{potential}
V&=& 4\,e^{2\,\varphi}\, e^{K}\, \left[\sum_{\varLambda=0}^3\,
(g_\varLambda)^2\,
|X^\varLambda|^2+\frac{1}{2}\,
(g_0^2+g_1^2)(t-\bar{t})\left((u-\bar{u})-\frac{1}{2}\,\frac{(
x^k-\bar{x}^k)^2}{(t-\bar{t})}\right)\right.\nonumber\\&&\left. +
\frac{(y^r-\bar{y}^r)^2}{8\,(s-\bar{s})(u-\bar{u})}\left(g_0^2\,(\bar{u}
x^k-\bar{x}^k u)^2+g_1^2\,(x^k-\bar{x}^k)^2\right) \right]\,.
\end{eqnarray}
From the above expression we see that in the ${\Scr N}=2$ case,
namely for $g_0=g_1=0$, the potential depends on $y^r$ only
through the factor $e^{K}$ and vanishes identically in $y^r$ for
the values of the $t,u$ scalars given in (\ref{n2tu}), for which
$X^2=X^3=0$. If $g_0$ or $g_1$ are non--vanishing (${\Scr
N}=1,\,0$ cases) the extremisation of the potential with respect
to $x^k$, namely $\partial_{x^k} V=0$ fixes $x^k=0$. For $x^k=0$
the potential depends on $y^r$ only through the factor $e^{K}$ and
vanishes identically in $y^r$ for $t=u={\rm i}$.

\subsection{Generalised gauging}

From a four--dimensional supergravity point of view we could
consider a generalisation of the previous gauging in which also
the D7 and D3--brane vectors $A^i_\mu,\,A^r_\mu$ are used to
gauge translational isometries along the directions $C^a$. We can
choose for simplicity to turn on couplings for all the $n_3$
D3--brane vectors and $n_7$ D7--brane vectors with the
constraint $n_3+n_7\le 17$.
These new couplings do not have an immediate interpretation in
terms of string compactification with fluxes. The constant Killing
vectors are $k^u_\varLambda=g^k_4$, $\varLambda=3+k,\, k=1,\dots,n_7$,
along the direction $q^u=C^{a=3,\dots,2+n_7}$ and
$k^u_\varLambda=g^r_5$, $\varLambda=3+n_7+r,\, r=1,\dots,n_3$, along the
direction $q^u=C^{a=3+n_7,\dots, 2+n_3+n_7}$. The expression of
the new potential, at the extremum of the $e^m_a$ scalars, is a
trivial extension of eq. (\ref{potential}):
\begin{eqnarray} 
\label{potential2}
V&=& 4\,e^{2\,\varphi}\, e^{K}\, \left[\sum_{\varLambda=0}^3\,
(g_\varLambda)^2\, |X^\varLambda|^2+\sum_{k=1}^{n_7} (g_4^k)^2\,
|X^{3+k}|^2+\sum_{r=1}^{n_3} (g_5^r)^2\,
|X^{3+n_7+r}|^2 \right.
\nonumber\\
&&\left. +
\frac{1}{2}\,(g_0^2+g_1^2)(t-\bar{t})\left((u-\bar{u})-\frac{1}{2}\,\frac{(
x^k-\bar{x}^k)^2}{(t-\bar{t})}\right) \right.
\nonumber\\
&&\left. +
\frac{(y^r-\bar{y}^r)^2}{8\,(s-\bar{s})(u-\bar{u})}\left(g_0^2\,(\bar{u}
x^k-\bar{x}^k u)^2+g_1^2\,(x^k-\bar{x}^k)^2\right) \right]\,.
 \end{eqnarray}
 As far as supersymmetric vacua are concerned, 
from inspection of the fermion shifts
 it is straightforward to verify that the existence of a
 constant Killing spinor always requires 
$X^{2},\,X^3,\,X^{3+k},\,X^{3+n_7+r}=0$
 which implies $x^k=y^r=0$ and $t=u={\rm i}$.
  As before we have ${\Scr N}=2$ if $g_0=g_1=0$,
 ${\Scr N}=1$ if $g_0=g_1\neq 0$ and ${\Scr N}=0$ otherwise. The ${\Scr N}=0$
 flat vacuum is also defined by the conditions $x=y=0$ and $t=u={\rm
 i}$, as it can be verified from eq. (\ref{potential2}). The
 presence of non--vanishing $g^k_4$ and $g^r_5$ couplings therefore
 fixes the positions of the branes along 
$T^2$ to $x^k=y^r=0$ in all the relevant cases.
\section{Conclusions}
The present investigation allows us to study in a fairly general
way the potential for the 3--form flux compactification, in
presence of both bulk and open string moduli. In absence of fluxes
the D3, D7 dependence of the K\"ahler potential is rather
different since this moduli couple in different ways to the bulk
moduli.

Moreover, in the presence of 3--form fluxes which break
${\Scr N}=2\rightarrow {\Scr N}=1,0$ the D7 moduli are
stabilised while the D3 moduli are not. For small values of the
coordinates $x^k$, $y^r$ the dependence of their kinetic term is
(for $u=t={\rm i}$), $-(\partial_\mu \bar{y}^r\partial^\mu
y^r)/{\rm Im}(s)$ for the D3--brane moduli, and $-(\partial_\mu
\bar{x}^k\partial^\mu x^k)$ for the D7--brane moduli. This is in
accordance with the suggestion of \cite{lateKallosh}. Note that
the above formulae, at $x=0,\,u=t={\rm i}$ are true up to
corrections $O(\frac{{Im}(y)^2}{{Im}(s)})$, since $y$ and $s$ are
moduli even in presence of fluxes. The actual dependence of these
terms on the compactification volume is important in order to
further consider models for inflatons where the terms in the
scalar potential allow to stabilise the remaining moduli.

Finally, we have not considered here the gauging of compact gauge groups
which exist on the brane world--volumes. This is, for instance, required 
\cite{dterm1,dterm2,KTW} in models with hybrid inflation \cite{hybrid}. 
This issue will be considered elsewhere.

\vskip 1.5truecm

\noindent{\bf Acknowledgements. }
We would like to thank Renata Kallosh for very useful discussions.
M.T. and R.D. would like to thank the Th. Division of CERN, where
part of this work has
 been done, for their kind hospitality.
Work supported in part by the European Community's Human Potential
Program under contract HPRN-CT-2000-00131 Quantum Space-Time, in
which  R. D'A. is associated to Torino University. The
work of S.F. has been supported in part by European Community's
Human Potential Program under contract HPRN-CT-2000-00131 Quantum
Space-Time, in association with INFN Frascati National
Laboratories and by D.O.E. grant DE-FG03-91ER40662, Task C.

\appendix
\section*{Appendix A \ \  Some relevant formulae. }\label{appendiceA}
\setcounter {equation}{0} \addtocounter{section}{1}
 
We are interested in gauging the 22
translations in the coset ${\rm SO}(4,20)/({\rm SO}(3,19)\times
{\rm O}(1,1))$. Let us denote by $L$ the coset representative of
${\rm SO}(3,19)/{\rm SO}(3)\times {\rm SO}(19)$. It will be
written in the form:
\begin{equation}
L = \left(\matrix{(1+{\bf e}\,{\bf e}^T)^{\frac{1}{2}} & -{\bf
e}\cr -{\bf e}^T & (1+{\bf e}^T{\bf e})^{\frac{1}{2}} }\right)
\end{equation}
where ${\bf e}=\{e^m{}_a\}$, ${\bf e}^T=\{e^a{}_m\}$ , $m=1,2,3$
and $a=1,\dots , 19$, are the coordinates of the manifold. The
$22$ nilpotent Peccei--Quinn generators are denoted by
$\{Z_m,\,Z_a\}$ and the gauge generators are:
\begin{equation}
t_\varLambda = f^m{}_\varLambda Z_m+ h^a{}_\varLambda Z_a
\end{equation}
the corresponding Killing vectors have non vanishing components:
$k^m_\varLambda=f^m{}_\varLambda$ and $k^a_\varLambda=h^a{}_\varLambda$. The
moment maps are:
\begin{equation}
{\Scr P}^x_\varLambda =
\sqrt{2}\,\left(e^{\varphi}\,(L^{-1})^x{}_m\,f^m{}_\varLambda
+e^{\varphi}\,(L^{-1})^x{}_a\,h^a{}_\varLambda\right)
\end{equation}
where $\varphi$ is the $T^2$ volume modulus \cite{ADFL}:
$e^{-2\,\varphi}={\rm Vol}(T^2)$ and $x=1,2,3$. The metric along
the Peccei--Quinn directions $I=(m,a)$ is:
\begin{equation}
h_{IJ}  = e^{2\,\varphi}\,(\delta_{IJ}+2\,e^a{}_I e^a{}_J)
\end{equation}
The potential has the following form:
\begin{eqnarray}
V&=&4\,
e^{2\,\varphi}\,\left(f^m{}_\varLambda\,f^m{}_\varSigma+2\,e^a{}_m
e^a{}_n\,f^m{}_\varLambda\,f^n{}_\varSigma+h^a{}_\varLambda\,
h^a{}_\varSigma\right)\,\bar{L}^\varLambda\,
L^\varSigma \nonumber\\
&& +
2\,e^{2\,\varphi}\,\left(U^{\varLambda\varSigma}-3\,\bar{L}^\varLambda\,
L^\varSigma\right)\,\left(f^m{}_\varLambda\,f^m{}_\varSigma+e^a{}_m
e^a{}_n\,f^m{}_\varLambda\,f^n{}_\varSigma \right.
\nonumber\\
&&\left. +
2\,[(1+{\bf e}\,{\bf e}^T)^{\frac{1}{2}}]^n{}_m e^n{}_a\,
f^m{}_{(\varLambda}\,h^a{}_{\varSigma)}+e^n{}_a
e^n{}_b\,h^a{}_\varLambda\,h^b{}_\varSigma\right) \,. \label{pot}
\end{eqnarray}
In all the models we consider, at the extremum point of the
potential in the special K\"ahler manifold the following condition
holds: $\left(U^{\varLambda\varSigma}-3\,\bar{L}^\varLambda\,
L^\varSigma\right)_{|0}\,f^m{}_{(\varLambda}\,h^a{}_{\varSigma)}=0$. 
As a consequence of this, as it is clear from (\ref{pot}), the
potential in this point depends on the metric scalars $e^m_a$ only
through quadratic terms in the combinations
$e^m{}_a\,h^a{}_\varLambda$ and $ e^a{}_m\,f^m{}_\varLambda$. Therefore
$V$ is extremised with respect to the $e^m_a$ scalars once we
restrict ourselves to the moduli defined as follows:
\begin{equation}
\mbox{moduli:} \qquad  e^m{}_a\,h^a{}_\varLambda =e^a{}_m\,f^m{}_\varLambda 
=0 \,.
\end{equation}
The vanishing of the potential implies
\begin{equation}
\left(U^{\varLambda\varSigma}-\,\bar{L}^\varLambda\,
L^\varSigma\right)_{|0}\,f^m{}_{(\varLambda}\,f^m{}_{\varSigma)}
+2 \left(\bar{L}^\varLambda\,
L^\varSigma\right)_{|0}\,h^a{}_{(\varLambda}\,h^a{}_{\varSigma)}=0 \,.
\end{equation}
Furthermore, one may notice that, as in \cite{ADFL}, the following
relations hold in all the models under consideration:
\begin{equation}
\left(U^{\varLambda\varSigma}-\,\bar{L}^\varLambda\,
L^\varSigma\right)_{|0}\,f^m{}_{(\varLambda}\,f^m{}_{\varSigma)}
=\left(\bar{L}^\varLambda\,
L^\varSigma\right)_{|0}\,h^a{}_{(\varLambda}\,h^a{}_{\varSigma)}=0 \,.
\end{equation}
Our analysis is limited to the case in which the only non--vanishing
$f$ and $h$ constants are:
\begin{equation}
f^1{}_0 = g_0 \,\,;\,\,\,f^2{}_1=g_1 \,\,;\,\,\,h^1{}_2=g_2
\,\,;\,\,\,h^2{}_3=g_3 \,\,;\,\,\,h^{2+k}{}_{3+k}=g^k_4
\,\,;\,\,\,h^{2+n_7+r}{}_{3+n_7+r}=g^r_5 \,.
\end{equation}

\end{document}